\documentstyle[12pt,a4]{article}

\begin{document}

\baselineskip=18pt plus 0.2pt minus 0.1pt

\makeatletter

\def\p{{\partial}}
\def\nn{{\nonumber}}
\newcommand{\be}{\begin{equation}}
\newcommand{\ee}{\end{equation}}
\newcommand{\bea}{\begin{eqnarray}}
\newcommand{\eea}{\end{eqnarray}}
\newcommand{\Tr}{\mathop{\rm Tr}}
\newcommand{\Pf}{\mathop{\rm Pf}}
\renewcommand{\thefootnote}{\fnsymbol{footnote}}

\begin{titlepage}
\title{
\hfill\parbox{4cm}
{\normalsize KUNS-1667\\{\tt hep-th/0006056}}\\
\vspace{1cm}
Noncommutative/Nonlinear BPS Equations\\
without Zero Slope Limit
}
\author{
Sanefumi {\sc Moriyama}
\thanks{{\tt moriyama@gauge.scphys.kyoto-u.ac.jp}}
\\[7pt]
{\it Department of Physics, Kyoto University, Kyoto 606-8502, Japan}
}
\date{\normalsize June, 2000}
\maketitle
\thispagestyle{empty}

\begin{abstract}
\normalsize
It is widely believed that via the Seiberg-Witten map, the linearly
realized BPS equation in the non-commutative space is related to the
non-linearly realized BPS equation in the commutative space in the
zero slope limit. We show that the relation also holds without taking
the zero slope limit as is expected from the arguments of the BPS
equation for the non-Abelian Born-Infeld theory. This is regarded as
an evidence for the relation between the two BPS equations. As a
byproduct of our analysis, the non-linear instanton equation is solved
exactly.
\end{abstract}

\end{titlepage}

\section{Introduction and summary}
Recently the string theory has been found to be more fertile than what 
it was thought previously. It contains many important concepts in
physics including, among other things, non-commutativity.
D-brane in the string theory with the background NS-NS 2-form $B_{ij}$
has two effective theories: the ordinary Born-Infeld theory when the
Pauli-Villars regularization is adopted \cite{ACNP} and the
non-commutative Born-Infeld theory when the point-splitting
regularization is adopted \cite{CDS,DH}. Since the method of
regularizations should not change the physical S-matrices, it was
discussed in \cite{SW} that these two descriptions should be related
by field redefinitions (called the Seiberg-Witten map).

This relation has also been explored from the classical solutions.
In fact, many solitons and instantons were constructed in the
non-commutative space \cite{NS,HasHatMor,Bak,HatMor,GMS,GH,GN} for
investigating many interesting properties by themselves
\cite{HasHas,BN,Fur,DMR,HKLM} and their relations to the commutative
space \cite{SW,Ter,Mat,HasHir,Mor}.
In the most cases except a few attempts \cite{Ter,Mor}
the relation was discussed in the zero slope limit.
In the non-commutative side the linearly realized BPS equation
$\hat F^+=0$ is directly obtained from the BPS bound of the Yang-Mills
theory \cite{HasHatMor} which is the zero slope limit of the
Born-Infeld theory.
Therefore it is widely believed that the linearly realized BPS
equation in the non-commutative space (which we abbreviate to the
non-commutative BPS equation) $\hat F^+=0$ is related to the
non-linearly realized BPS equation in the commutative space
(abbreviated to the non-linear BPS equation) in the zero slope limit
$(F+B)^+/\Pf(F+B)=B^+/\Pf B$ by the Seiberg-Witten map.\footnote{
The non-linear BPS equation in the commutative space may be subject to 
derivative corrections of the string effective action. We shall
discuss this issue in the final section.}
This was implicitly pointed out in \cite{SW} by rewriting the
non-linear BPS equation in the zero slope limit in terms of the open
string moduli: the open string metric $G_{ij}$ and the
non-commutativity parameter $\theta^{ij}$, which originally appears in
the non-commutative BPS equation and the Seiberg-Witten map.
Though persuasive, it is still difficult to show the relation
explicitly because the method used in \cite{SW} to connect two
Born-Infeld actions cannot be applied to the case of the BPS
equations.

On the other hand, it was shown in \cite{AHas,Bre} that although the
linear BPS equation is obtained directly from the Yang-Mills theory,
it also reproduces the equation of motion of the Born-Infeld theory if
we adopt the symmetrized trace prescription \cite{Tse}.
Though this was shown in the non-Abelian case, we can generalize the
argument to the non-commutative case straightforwardly.
The fact that the non-commutative BPS equation is unchanged in the
zero slope limit implies that its commutative counterpart also remains 
the same under the zero slope limit because the Seiberg-Witten map
$
\delta\hat A_i=-1/4\delta\theta^{kl}
\Bigl(\hat A_k*(\p_l\hat A_i+\hat F_{li})
+(\p_l\hat A_i+\hat F_{li})*\hat A_k\Bigr)
$
does not depend on the slope $\alpha'$.

In this paper we rewrite the non-linear BPS equation in terms of the
open string moduli without taking the zero slope limit.
We show that, even though we do not take the zero slope limit, the
non-linear BPS equation in the open string moduli takes completely the
same form as that obtained in the zero slope limit.
This calculation is also motivated by our previous work \cite{Mor}
where the non-linear BPS monopole solution without taking the zero
slope limit was rewritten in terms of the open string
moduli.\footnote{
In the monopole case \cite{Mor} we only turned on the spatial
$B$-field background. This corresponds to the special case of
$\Pf B=0$ in the instanton terminology and apparently leads to a
singular BPS equation if we take the zero slope limit. That is why we
tried to discuss without taking the zero slope limit in \cite{Mor}.}

This result shows that taking the zero slope limit is unnecessary in
discussing the relation between the non-commutative BPS equation and
non-linear BPS equation.
More importantly, this result gives us a strong evidence for the
conjecture that the non-commutative BPS equation is related to the
non-linear BPS equation, because the invariant nature of the
non-commutative BPS equation under the zero slope limit is perfectly
reproduced in the commutative side.\footnote{
Another interesting evidence is that both in the non-commutative
side \cite{Bak,GN} and the commutative side \cite{Mor} the monopole
moduli space is unchanged under the deformation of the
non-commutativity parameter.
}

In the next section, we shall explicitly rewrite the non-linear BPS
equation in terms of the open string moduli without taking the zero
slope limit. And we shall discuss the physical implications and their
applications in the final section.

\section{Nonlinear BPS equation in the open string moduli}
In this section we shall rewrite the non-linear BPS equation in terms
of the open string moduli.
First let us recall the linear supersymmetries and non-linear
supersymmetries of gauginos \cite{BG,Ket,Tse}:
\bea
\delta_{\rm L}\lambda_+
&=&\frac{1}{2\pi\alpha'}M_{ij}^+\sigma^{ij}\eta,\\
\delta_{\rm L}\lambda_-
&=&\frac{1}{2\pi\alpha'}M_{ij}^-\bar\sigma^{ij}\bar\eta,\\
\delta_{\rm NL}\lambda_+&=&\frac{1}{4\pi\alpha'}
\Bigl(1-\Pf M+\sqrt{1-\Tr M^2/2+(\Pf M)^2}\Bigr)\eta^*,\\
\delta_{\rm NL}\lambda_-&=&\frac{1}{4\pi\alpha'}
\Bigl(1+\Pf M+\sqrt{1-\Tr M^2/2+(\Pf M)^2}\Bigr)\bar\eta^*,
\eea
where $M$ denotes
\bea
M=2\pi\alpha'(F+B),
\eea
with the field strength $F_{ij}$ and the background NS-NS 2-form
$B_{ij}$. Hereafter we shall set $2\pi\alpha'=1$ for simplicity,
however we can restore it on the dimensional ground.
At the infinity the field strength vanishes and the combination of the 
unbroken supersymmetries is given as
\bea
B_{ij}^+\sigma^{ij}\eta
+\frac12\Bigl(1-\Pf B+\sqrt{1-\Tr B^2/2+(\Pf B)^2}\Bigr)\eta^*=0.
\eea
The non-linear BPS equation is the condition of preserving these
supersymmetries:
\bea
\frac{M^+}{1-\Pf M+\sqrt{1-\Tr M^2/2+(\Pf M)^2}}
=\frac{B^+}{1-\Pf B+\sqrt{1-\Tr B^2/2+(\Pf B)^2}}.
\label{BPS}
\eea

For rewriting this non-linear BPS equation (\ref{BPS}) in terms of the
open string moduli, we shall first rewrite it into a simpler form.
First note from eq.\ (\ref{BPS}), the matrix $M^+$ must be
proportional to $B^+$:
\bea
M^+=fB^+.
\label{proportion}
\eea
Rewriting eq.\ (\ref{BPS}) into a scalar equation as
\bea
f\Bigl(1-\Pf B+\sqrt{1-\Tr B^2/2+(\Pf B)^2}\Bigr)-(1-\Pf M)
=\sqrt{1-\Tr M^2/2+(\Pf M)^2},
\label{scalareq}
\eea
by using eq.\ (\ref{proportion}) and taking the square of eq.\
(\ref{scalareq}), eq.\ (\ref{BPS}) is reduced to a much 
simpler form \cite{MMMS,Ter}:
\bea
\frac{M^+}{1-\Pf M}=\frac{B^+}{1-\Pf B}.
\label{reducedBPS}
\eea
Here we have used the following identities,
\bea
\Tr M^2&=&\Tr(M^+)^2+\Tr(M^-)^2,\\
4\Pf M&=&-\Tr(M^+)^2+\Tr(M^-)^2.
\eea
Note that if we further use the identity,
\bea
\Pf(F+B)=\Pf F+\Pf B-\Tr F\tilde B/2,
\eea
eq.\ (\ref{reducedBPS}) now reads
\bea
F^+(1-\Pf B)=B^+(\Tr F\tilde B/2-\Pf F).
\label{closed}
\eea

Now let us proceed to rewriting eq.\ (\ref{closed}) in terms of the
open string moduli: the open string metric $G_{ij}$ and the
non-commutativity parameter $\theta^{ij}$.
The open string moduli is related to the closed string
moduli as \cite{SW}
\bea
\frac{1}{G}+\theta=\frac{1}{g+B}.
\eea
Since we adopt the flat metric for the closed string metric
$g_{ij}=\delta_{ij}$, the open string metric $G_{ij}$ and the
non-commutativity parameter $\theta^{ij}$ are expressed in terms of
the $B$-field:
\bea
G_{ij}&=&\delta_{ij}-(B^2)_{ij},
\label{metric}\\
\theta^{ij}&=&\frac{-B_{ij}-\tilde B_{ij}\Pf B}{\det(1+B)}.
\label{noncomm}
\eea
Since in eq.\ (\ref{closed}) the self-dual projection appears, we also
expect it to appear in the BPS equation in the open string moduli.
As we know from \cite{SW} the easiest way to write down the self-dual
projection is neither in the covariant frame nor in the contravariant
one but in the local Lorentz frame.
Hence we have to calculate
\bea
\underline{F}^+&=&(E^tFE)^+,\\
\underline{\theta}^+&=&\biggl(\frac{1}{E}\theta\frac{1}{E^t}\biggr)^+.
\eea
Here the vierbein is defined as $EGE^t=1$.
From the metric (\ref{metric}) we find that the vierbein is given as
\bea
E=(1+B)^{-1}.
\eea

In calculating the self-dual projection of the field strength,
we shall go to a special frame where $B$ has the canonical form as in
\cite{SW}:
\bea
B=\pmatrix{0&b_1&0&0\cr -b_1&0&0&0\cr 0&0&0&b_2\cr 0&0&-b_2&0}.
\eea
In this frame $\underline{F}^+$ is given as
\bea
\underline{F}^+
=\biggl(\frac{1}{1-B}F\frac{1}{1+B}\biggr)^+
=\frac{1}{(1+b_1^2)(1+b_2^2)}
\pmatrix{0&f_1&f_2&f_3\cr -f_1&0&f_3&-f_2\cr
-f_2&-f_3&0&f_1\cr -f_3&f_2&-f_1&0},
\eea
where $f_1$, $f_2$ and $f_3$ denote
\bea
2f_1&=&(1+b_2^2)F_{12}+(1+b_1^2)F_{34},\\
2f_2&=&(1-b_1b_2)(F_{13}-F_{24})+(b_1+b_2)(F_{14}+F_{23}),\\
2f_3&=&(1-b_1b_2)(F_{14}+F_{23})-(b_1+b_2)(F_{13}+F_{24}).
\eea
We can easily identify terms proportional to $F^+$ and $B^+$, however
there are still other terms to be identified as $B^+F^+-F^+B^+$:
\bea
\biggl(\frac{1}{1-B}F\frac{1}{1+B}\biggr)^+
=\frac{(1-\Pf B)F^+-(\Tr F\tilde B)B^+/2+(B^+F^+-F^+B^+)}{\det(1+B)}.
\eea

On the other hand, $\underline{\theta}^+$ is much easier to
calculate. We find
\bea
\underline{\theta}^+=\Bigl((1+B)\theta(1-B)\Bigr)^+=-B^+,
\eea
where we have used
\bea
B^3&=&(\Tr B^2)B/2+(\Pf B)\tilde B,\\
B\tilde BB&=&-(\Pf B)B,
\eea
which hold for any anti-symmetric 4$\times$4 matrix $B$.

Since we are considering the non-linear BPS equation (\ref{BPS})
which implies that $F^+$ is proportional to $B^+$ (\ref{proportion}),
it is possible to add $(B^+F^+-F^+B^+)$ to eq.\ (\ref{closed})
freely because we have $B^+F^+-F^+B^+=0$.
Collecting all of our results and comparing them with eq.\
(\ref{closed}) we find it remains to rewrite $\Pf F$ into the local
Lorentz frame:
\bea
\Pf\underline{F}=\frac{\Pf F}{\det(1+B)}.
\eea
Therefore our non-linear BPS equation (\ref{BPS}) is finally rewritten
as
\bea
\underline{F}^+=\underline{\theta}^+\Pf\underline{F}.
\label{open}
\eea
Amazingly, this is the same form as that obtained in the zero slope
limit \cite{SW}.

\section{Physical implications and further directions}
First, our analysis in this paper is important in the conceptual
sense. We have given another strong evidence for the fact that the
non-commutative BPS equation is mapped to the non-linear BPS equation,
because the invariant nature of the non-commutative BPS equation under
the zero slope limit is perfectly reproduced in the commutative side.
In fact, the non-linear BPS equation without taking the zero slope
limit has the same form as that obtained in the zero slope limit
\cite{SW} when they are rewritten in terms of the open string moduli
(\ref{open}).

Secondly, so far we have neglected the stringy derivative corrections
to the Born-Infeld theory.
Strictly speaking, it is the non-commutative and commutative
Born-Infeld theory with derivative corrections that should be related
by the Seiberg-Witten map, not the Born-Infeld theories themselves.
Hence if we would like to discuss the relation between the two BPS
equations, we have to see if the two BPS equations remain unchanged
when the derivative corrections are taken into account.
In the commutative side, the solutions of the linear BPS equation are
also solutions of the Born-Infeld theory with corrections
\cite{Tho}.
However we do not know a similar argument for the non-linear BPS
equation. To clarify this point is an interesting direction.
Our result (\ref{open}) might give a clue to this question.

Thirdly, as a technical application this rewriting enables us to find
an instanton solution to the non-linear BPS equation for a general
constant $B$-field background without taking the zero slope limit.
In \cite{Ter} the solution is constructed under the condition $B^-=0$,
because otherwise the solution is very intricate.
However, the non-linear BPS equation is now rewritten as (\ref{open}),
which is completely solved in \cite{SW}. Hence it is possible to read
off the solution to the non-linear BPS equation directly:
\bea
A_i=\theta^+_{ij}x^j
\cdot\frac{1}{4}
\Biggl(-1+\sqrt{1+\frac{32C}{R^4}}\Biggr),
\label{B^-}
\eea
with $2\theta^+_{ij}=-(1-\Tr B^2/2+\Pf B)B_{ij}-(1-\Pf B)\tilde B_{ij}$
and $R^2=x^i(1-B^2)_{ij}x^j$.
Though the solution should be independent of $\theta^-$ as noted in
\cite{SW}, our solution (\ref{B^-}) has a non-trivial dependence on
$B^-$.
In this way the commutative counterparts of the non-commutative Abelian
instanton and monopole \cite{NS,GN} with or without taking the zero
slope limit are all constructed \cite{SW,Mor}.

Finally, since we have rewritten the non-linear BPS equation
completely in terms of the open string moduli, we should expect the
tension of the exact monopole solution found in \cite{Mor} also has a
trivial dependence on $\alpha'$.
However, it was discussed in \cite{GN} that this is not the case and
there might be a discrepancy between the non-commutative and
commutative viewpoints.
It is very important to resolve this discrepancy.

{\bf Note added}\\
After submitting this paper for the publication, the authors of
\cite{GN} made a revision to resolve the discrepancy between the
monopole tensions mentioned in Sec.\ 3 of this paper and showed that
they do not depend on $\alpha'$.

{\bf Acknowledgement}\\
We would like to thank K.\ Hashimoto, H.\ Hata, T.\ Noguchi and S.\
Terashima for valuable discussions and comments.
The author is also grateful to H.\ Hata for his encouragement and
careful reading of the present manuscript.
This work is supported in part by Grant-in-Aid for Scientific Research
from Ministry of Education, Science, Sports and Culture of Japan
(\#04633). The author is supported in part by the Japan Society for the
Promotion of Science under the Predoctoral Research Program.

\newcommand{\J}[4]{{\sl #1} {\bf #2} (#3) #4}
\newcommand{\AP}{Ann.\ Phys.\ (N.Y.)}
\newcommand{\MPL}{Mod.\ Phys.\ Lett.}
\newcommand{\NP}{Nucl.\ Phys.}
\newcommand{\PL}{Phys.\ Lett.}
\newcommand{\PR}{Phys.\ Rev.}
\newcommand{\PRL}{Phys.\ Rev.\ Lett.}
\newcommand{\PTP}{Prog.\ Theor.\ Phys.}
\newcommand{\hep}[1]{{\tt hep-th/{#1}}}


\begin{thebibliography}{99}
\bibitem{ACNP}
A.~Abouelsaood, C.~G.~Callan, C.~R.~Nappi and S.~A.~Yost,
``Open Strings in Background Gauge Fields'',
\J{\NP}{B280}{1987}{599}.

\bibitem{CDS}
A.~Connes, M.~R.~Douglas and A.~Schwarz,
``Noncommutative Geometry and Matrix Theory:
Compactification on Tori'',
\J{JHEP}{02}{1998}{003}, {\tt hep-th/9711162}.

\bibitem{DH}
M.~R. Douglas and C.~Hull, ``D-branes and the Noncommutative Torus'',
\J{JHEP}{02}{1998}{008}, {\tt hep-th/9711165}.

\bibitem{SW}
N.~Seiberg and E.~Witten,
``String Theory and Noncommutative Geometry'',
\J{JHEP}{09}{1999}{032}, {\tt hep-th/9908142}.

\bibitem{NS}
N.~Nekrasov and A.~Schwarz,
``Instantons on noncommutative $R^4$,
and (2,0) superconformal six dimensional theory'',
\J{Commun.\ Math.\ Phys.}{198}{1998}{689}, {\tt hep-th/9802068}.

\bibitem{HasHatMor}
K.~Hashimoto, H.~Hata and S.~Moriyama,
``Brane Configuration from Monopole Solution
in Non-Commutative Super Yang-Mills Theory'',
\J{JHEP}{12}{1999}{021}, {\tt hep-th/9910196}.

\bibitem{Bak}
D.~Bak, ``Deformed Nahm Equation and a Noncommutative BPS Monopole'',
\J{\PL}{B471}{1999}{149}, {\tt hep-th/9910135}.

\bibitem{HatMor}
H.~Hata and S.~Moriyama,
``String Junction from Non-Commutative Super Yang-Mills Theory'',
\J{JHEP}{02}{2000}{011}, {\tt hep-th/0001135}.

\bibitem{GMS}
R.~Gopakumar, S.~Minwalla and A.~Strominger,
``Noncommutative Solitons'',
\J{JHEP}{05}{2000}{020}, {\tt hep-th/0003160}.

\bibitem{GH}
S.~Goto and H.~Hata,
``Noncommutative Monopole at the Second Order in $\theta$'',
{\tt hep-th/0005101}.

\bibitem{GN}
D.~J.~Gross and N.~A.~Nekrasov,
``Monopoles and Strings in Noncommutative Gauge Theory'',
{\tt hep-th/0005204}.

\bibitem{HasHas}
A.~Hashimoto and K.~Hashimoto,
``Monopoles and Dyons in Non-Commutative Geometry'',
\J{JHEP}{11}{1999}{005}, {\tt hep-th/9909202}.

\bibitem{BN}
H.~W.~Braden and N.~A.~Nekrasov,
``Space-Time Foam From Non-Commutative Instantons'',
{\tt hep-th/9912019}.

\bibitem{Fur}
K.~Furuuchi,
``Instantons on Noncommutative $R^4$ and Projection Operators'',\\
{\tt hep-th/9912047}.

\bibitem{DMR}
K.~Dasgupta, S.~Mukhi and G.~Rajesh, ``Noncommutative Tachyons'',
{\tt hep-th/0005006}.

\bibitem{HKLM}
J.~A.~Harvey, P.~Kraus, F.~Larsen and E.~J.~Martinec,
``D-branes and Strings as Non-commutative Solitons'',
{\tt hep-th/0005031}.

\bibitem{Ter}
S.~Terashima, 
``U(1) Instanton in Born-Infeld Action
and Noncommutative Gauge Theory'',
\J{\PL}{B477}{2000}{292}, {\tt hep-th/9911245}.

\bibitem{Mat}
D.~Mateos,
``Non-commutative vs. Commutative Descriptions of D-brane BIons'',
{\tt hep-th/0002020}.

\bibitem{HasHir}
K.~Hashimoto and T.~Hirayama,
``Branes and BPS Configurations of Non-Commutative 
/Commutative Gauge Theories'',
{\tt hep-th/0002090}.

\bibitem{Mor}
S.~Moriyama, ``Noncommutative Monopole from Nonlinear Monopole'',
{\tt hep-th/0003231}, to appear in {\sl \PL} {\bf B}.

\bibitem{AHas}
A.~Hashimoto, ``The Shape of Branes Pulled by Strings'',
\J{\PR}{D57}{1998}{6441}, {\tt hep-th/9711097}.

\bibitem{Bre}
D.~Brecher, ``BPS States of the Non-Abelian Born-Infeld Action'',
\J{\PL}{B442}{1998}{117}, {\tt hep-th/9804180}.

\bibitem{Tse}
A.~A.~Tseytlin,
``Born-Infeld action, supersymmetry and string theory'',
{\tt hep-th/9908105}.

\bibitem{BG}
J.~Bagger and A.~Galperin,
``New Goldstone multiplet for partially broken supersymmetry'',
\J{\PR}{D55}{1997}{1091}, {\tt hep-th/9608177}.

\bibitem{Ket}
S.~V.~Ketov,
``A manifestly $N=2$ supersymmetric Born-Infeld action'',
\J{\MPL}{A14}{1999}{501}, {\tt hep-th/9809121}.

\bibitem{MMMS}
M.~Marino, R.~Minasian, G.~Moore and A.~Strominger, 
``Nonlinear Instantons from Supersymmetric $p$-Branes'',
\J{JHEP}{01}{2000}{005}, {\tt hep-th/9911206}.

\bibitem{Tho}
L.~Thorlacius,
``Born-Infeld String as a Boundary Conformal Field Theory'',
\J{\PRL}{80}{1998}{1588}, {\tt hep-th/9710181}.

\end{thebibliography}
\end{document}